\documentclass[conference]{IEEEtran}
\IEEEoverridecommandlockouts

\usepackage{amsmath,amssymb,amsfonts}
\usepackage{algorithmic}
\usepackage{graphicx}
\usepackage{textcomp}
\usepackage{enumitem}
\usepackage{float}
\usepackage{xcolor}
\usepackage[dvipsnames]{xcolor}
\usepackage{multirow}
\usepackage{orcidlink}
\usepackage[numbers,sort&compress]{natbib}
\def\BibTeX{{\rm B\kern-.05em{\sc i\kern-.025em b}\kern-.08em
    T\kern-.1667em\lower.7ex\hbox{E}\kern-.125emX}}
\begin{document}

\title{Speech Language Models for Full-Meeting Speaker Diarization: Capabilities and Limitations}

\author{
\IEEEauthorblockN{
Jialu Li\textsuperscript{1}\orcidlink{0000-0003-0092-8071}\thanks{Part of this work was conducted while Jialu Li was visiting the Carnegie Mellon University.},
Jinchuan Tian\textsuperscript{2}\orcidlink{0000-0002-2129-471X},
Shinji Watanabe\textsuperscript{2}\orcidlink{0000-0002-5970-8631}
}
\IEEEauthorblockA{
\textsuperscript{1}\textit{College of Information Science, University of Arizona},
Tucson, AZ, USA \\
\textsuperscript{2}\textit{Language Technologies Institute, Carnegie Mellon University},
Pittsburgh, PA, USA \\
}
}


\maketitle

\begin{abstract}
Recent advances in Speech Language Models (SpeechLMs), which integrate large language models with speech foundation models, have enabled unified sequence modeling of speech processing tasks. However, many SpeechLM-based approaches to speaker diarization (SD) are tightly coupled with automatic speech recognition (ASR) and evaluated using word-level metrics, making it difficult to assess SD performance independent of ASR accuracy. In this work, we investigate ESPnet-SpeechLM as a token-based backbone for generating SD hypotheses, formulating SD as autoregressive generation of structured tokens conditioned on acoustic input. We systematically compare two output representations: an event-based representation that explicitly models speaker turn onset and offset timestamps, and a frame-based representation that predicts frame-level speaker activity. To provide structured conversational cues, we further incorporate auxiliary tasks including speech activity detection, overlapped speech detection, and speaker turn counting within the output sequence. Across multiple meeting datasets, we find that event-based representations produce more stable and consistent SD outputs than frame-based representations. Our analysis shows that outputs generated by SpeechLMs encode useful temporal SD structure, but full-meeting SD remains limited by recording-level speaker tracking and overlap-related misses. Explicit speaker-linking post-processing substantially reduces speaker confusion, suggesting that robust SpeechLM-based SD requires persistent speaker tracking and overlap-aware generation.

\end{abstract}


\section{Introduction}
Speaker diarization (SD), determining “who spoke when”, is essential for downstream tasks, such as meeting transcription and conversational analysis. Full-meeting diarization benchmarks such as AMI~\cite{kraaij2005ami}, CHiME-6~\cite{watanabe20b_chime}, and NOTSOFAR-1~\cite{vinnikov2024notsofar} further highlight the practical importance of SD in multi-speaker, often far-field conversational settings. Traditional pipelines rely on multiple modules, including speech activity detection, segmentation, speaker embedding extraction, and clustering~\cite{park2022review}, while end-to-end neural diarization (EEND)~\cite{fujita19_interspeech} jointly learns speaker activity and identity, often using large-scale simulated mixtures. Recent work further investigates self-supervised learning (SSL) models to improve SD performance~\cite{han2025leveraging}. However, these systems are explicitly designed for SD and do not naturally extend to unified sequence modeling with other speech processing tasks.

Recent advances in Speech Language Models (SpeechLMs), which combine speech foundation models with large language models (LLMs), have enabled unified sequence modeling of acoustic and linguistic information~\cite{aroralandscape}. This has motivated LLM- and SpeechLM-based speaker-aware speech processing, including LLM-based speaker-label refinement~\cite{wang24h_interspeech,10446204} and token-based generation of transcripts, timestamps, and speaker labels for speaker-attributed transcription or short-region SD~\cite{li23o_interspeech,cornell2024one,huo2026tagspeech,yin2026speakerlm,yin2026whisperdiari,zheng2025dncasr}. Recent full-meeting systems further address cross-chunk speaker consistency using speaker-cache or cache-conditioned tracking mechanisms~\cite{shi2026train,peng2026gstar}. However, most of these systems are ASR-coupled, evaluated with word-level speaker-attribution metrics, or rely on explicit speaker-tracking mechanisms. In contrast, our work isolates SD from ASR and evaluates SpeechLM-generated SD under full-meeting DER.

Together, these studies show the promise of token-based speaker-aware modeling, but they often conflate SD with ASR. Speaker-attributed transcription combines word recognition, alignment, local speaker assignment, and cross-chunk speaker consistency, so word-level metrics do not directly isolate SD performance. In addition, many SpeechLM-style systems evaluate SD only on short utterance groups or fixed-length segments~\cite{li23o_interspeech,yin2026whisperdiari}. 
Full-meeting SD instead requires precise activity prediction and consistent recording-level speaker identities across multiple segments.
\begin{figure*}[t] 
    \centering
    \includegraphics[width=1.0\linewidth]{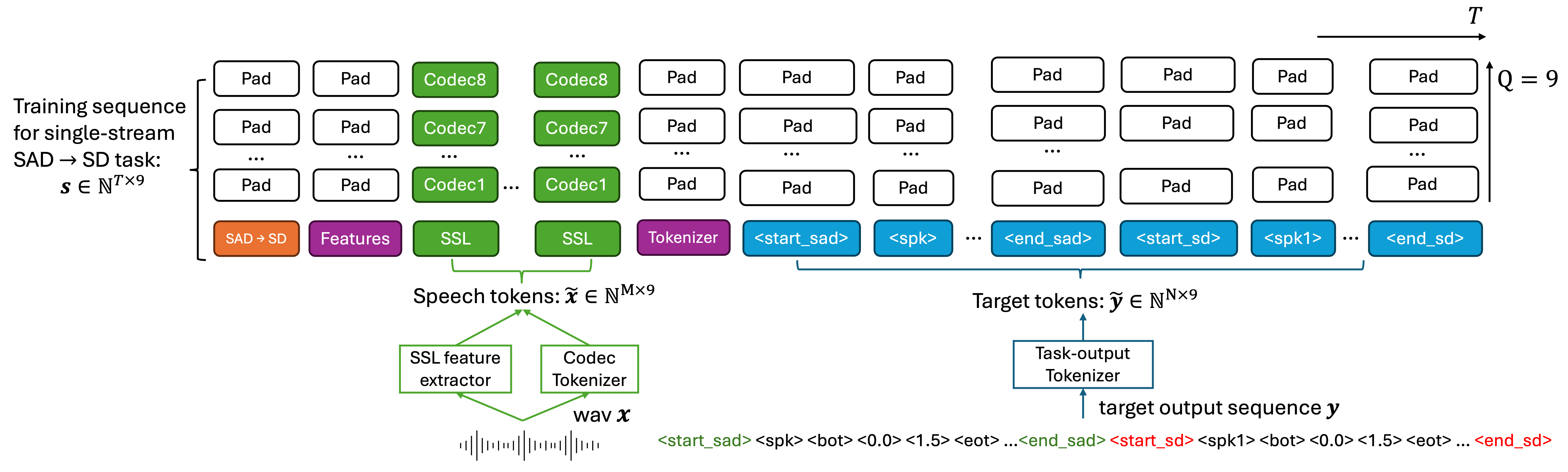} 
    \vspace{-0.6cm}
\caption{
Overview of the ESPnet-SpeechLM training sequence with single-stream task-output organization.
The input waveform is converted into speech tokens
$\tilde{\mathbf{x}} \in \mathbb{N}^{M \times 9}$,
where each frame contains eight \textcolor{ForestGreen}{codec} tokens
and one \textcolor{ForestGreen}{SSL} token.
A task marker (\textcolor{BurntOrange}{SAD$\rightarrow$SD}) and
modality-indicator tokens
(\textcolor{Fuchsia}{Features}, \textcolor{Fuchsia}{Tokenizer})
mark the transition from the speech-token sequence to the target-output sequence
$\tilde{\mathbf{y}} \in \mathbb{N}^{N \times 9}$.
The \textcolor{Cyan}{SAD} and \textcolor{Cyan}{SD} outputs are serialized
sequentially in a single target stream, while the remaining streams are zero-padded.
}
    \vspace{-0.5cm}
    \label{fig:model}
\end{figure*}

In this work, we use full-meeting SD as a diagnostic task for analyzing the speaker-temporal modeling capabilities of SpeechLMs. We formulate SD as autoregressive generation of structured tokens conditioned on acoustic input, and study how output format and auxiliary conversational cues affect this formulation by comparing \texttt{event}-based representations, which encode speaker turns as timestamped tuples, and \texttt{frame}-based representations, which emit speaker-activity tokens at fixed time steps. We further incorporate speech activity detection (SAD), overlapped speech detection (OD), and speaker turn counting (STC) within the output sequence. These tasks provide complementary cues for SD: coarse speech activity, overlap structure, and turn-taking information. Although auxiliary tasks such as SAD and OD have been explored in EEND-based models~\cite{takashima2021end}, their integration into SpeechLM-style token-based SD has not been previously examined.
Our main contributions are as follows:

\begin{itemize}[leftmargin=*, nosep] 
\item We formulate SD as autoregressive generation problem and compare \texttt{event}- and \texttt{frame}-based output representations under diarization error rate (DER) evaluation, decoupled from word-level ASR metrics.
\item We show that \texttt{event}-based outputs are more compact and more stable for meeting recordings, while \texttt{frame}-based generation becomes brittle as output sequences grow longer. Arranging SAD and OD before the final SD task in a coarse-to-fine manner consistently improves SD performance.
\item We provide a speaker-linking analysis that separates temporal SD structure from recording-level speaker tracking. Our results show that raw SpeechLM speaker symbols behave as local speaker hypotheses rather than recording-level identities, while explicit speaker linking substantially reduces speaker confusion.
\end{itemize}

\begin{table}[t]
\centering
\setlength{\tabcolsep}{2pt}
\caption{Statistics of datasets used in training, development, and testing set. }
\resizebox{\columnwidth}{!}{
\begin{tabular}{l|ccc|ccc|ccc}
\hline
\multirow{2}{*}{Dataset} & \multicolumn{3}{c|}{Train} & \multicolumn{3}{c|}{Dev} & \multicolumn{3}{c}{Test} \\
 & \#files & \#spk & \#hrs & \#files & \#spk & \#hrs & \#files & \#spk & \#hrs \\
\hline
AMI & 134 & 3-5 & 79.7 & 18 & 4 & 9.7 & 16 & 3-4 & 9.1 \\
AISHELL-4 & 173 & 3-5 & 89.8 & 18 & 3-5 & 9.3 & 20 & 5 & 9.2 \\
AliMeeting & 209 & 2-4 & 111.4 & 8 & 2-4 & 4.2 & 20 & 2-4 & 10.8 \\
\hline
Simulated & 40,000 & 2-5 & 3303.8 & 800 & 2-5 & 65.9 & 800 & 2-5 & 65.9 \\
\hline
\end{tabular}
}
\vspace{-0.3cm}
\label{tab:dataset_stats}
\end{table}

\section{Data}
Table~\ref{tab:dataset_stats} summarizes the dataset statistics for the training, development, and test partitions. We primarily investigate our SpeechLM-based SD system on  AMI~\cite{Renals2007}. We train and test on the AMI IHM-Mix recordings, which mixes all individual headset microphone signals. 
To evaluate the generalization and robustness, we further train and test on two additional Mandarin conversational corpora: AISHELL-4~\cite{AISHELL-4_2021} and AliMeeting~\cite{Yu2022Summary}.
AISHELL-4 includes recordings with up to seven speakers. 
Because our task-output vocabulary uses explicit speaker-symbol tokens \texttt{<spk1>}--\texttt{<spk5>}, we restrict each recording to a fixed five-speaker inventory to keep the speaker and overlap-label vocabulary tractable. Approximately 10\% of the data involve speakers beyond this inventory and are therefore affected by this restriction. We assign recording-level speaker IDs by first appearance and retain only segments involving speakers within this inventory. 
For AliMeeting, we use recordings captured by far-field microphones.
For all three datasets, we follow the training, development, and test splits described in~\cite{han2025leveraging}.

We further adapt our model using a large-scale simulated dataset.  
We follow the simulation recipe of~\cite{Fujita2019}, using Switchboard-2 (Phase I, II, III)~\cite{graff1998switchboard2,graff1999switchboard2phase2,graff2002switchboard2phase3}, Switchboard Cellular (Part 1,
Part2)~\cite{graff2001switchboardcellular1,graff2004switchboardcellular2}, and NIST Speaker Recognition Evaluation datasets (2004,
2005, 2006, 2008)~\cite{doddington2000nist} to generate 3303.8 hours of simulated conversations with two to five speakers and overlap ratios matched to real meeting datasets. 

\section{Methods}

\subsection{Model Architecture}
\label{sec:model_arch}
We use the open-source ESPnet-SpeechLM~\cite{watanabe2018espnet,tian2025espnet}, which formulates speech processing tasks as autoregressive sequence modeling with a decoder-only Transformer. 
ESPnet-SpeechLM represents speech and target outputs using a nine-stream token format. Figure~\ref{fig:model} illustrates this format under the single-stream organization used for event-based outputs.
Given an input wavform $\mathbf{x}$ and a target output sequence $\mathbf{y}$, the model tokenizes them into discrete speech tokens $\tilde{\mathbf{x}}$ and target tokens $\tilde{\mathbf{y}}$. The speech tokens are represented as $\tilde{\mathbf{x}} \in \mathbb{N}^{M \times 9}$, where $M$ is the number of speech frames. Each frame contains eight codec tokens and one SSL token extracted from XEUS~\cite{chen2024towards} using a 25\,ms window and a 10\,ms frame shift.
The target tokens are represented as $\tilde{\mathbf{y}} \in \mathbb{N}^{N \times 9}$, where $N$ is the number of output decoding steps. They are aligned and padded to match the multi-stream token format of the speech representation. We tokenize $\mathbf{y}$ using a compact task-output vocabulary that includes speaker IDs, timestamps, speaker-turn transition tokens, and auxiliary-task markers. The output vocabulary supports up to five speakers and timestamp tokens covering each 30-second segment. In addition to task-output tokens, modality-indicator tokens, such as \textcolor{Fuchsia}{Features} and \textcolor{Fuchsia}{Tokenizer} in Figure~\ref{fig:model}, are used to mark transitions between the speech-token stream and the target-output stream.

During training, the speech and target tokens are concatenated into a single sequence $\mathbf{s} = [\tilde{\mathbf{x}}, \tilde{\mathbf{y}}]$. We use teacher forcing and optimize cross-entropy loss over target tokens predicted from $p(\tilde{y}_{n,q} \mid \tilde{\mathbf{x}}, \tilde{\mathbf{y}}_{<n})$, where $\tilde{\mathbf{y}}_{<n}$ denotes the target-token vectors before output step $n$. At inference, $\tilde{\mathbf{y}}_{<n}$ is replaced by previously generated target-token vectors as the model autoregressively generates $\tilde{\mathbf{y}}$, which is detokenized into the final SD output.
The training loss is computed over valid target-token positions in the multi-stream sequence, excluding zero-padded positions. Let $\tilde{y}_{n,q}$ denote the target token at output step $n$ and stream $q$, with $q=1,\ldots,9$, and let $\Omega$ denote the set of valid non-padded target-token positions. The loss is defined as

\vspace{-0.2cm}
\begin{equation}
\label{eq}
\small
\begin{aligned}
\mathcal{L}
= - \sum_{(n,q)\in\Omega} \Big[
&\lambda_{\mathrm{mod}} \mathbb{I}(\tilde{y}_{n,q} \in \mathcal{V}_{\mathrm{mod}})
\log p(\tilde{y}_{n,q} \mid \tilde{\mathbf{x}}, \tilde{\mathbf{y}}_{<n}) \\
&+ \lambda_{\mathrm{out}} \mathbb{I}(\tilde{y}_{n,q} \in \mathcal{V}_{\mathrm{out}})
\log p(\tilde{y}_{n,q} \mid \tilde{\mathbf{x}}, \tilde{\mathbf{y}}_{<n})
\Big].
\end{aligned}
\end{equation}

where $n$ indexes autoregressive output steps, $q$ indexes the nine token streams, and $\tilde{y}_{n,q}$ is the ground-truth target token at step $n$ and stream $q$. $\mathcal{V}_{\text{mod}}$ and $\mathcal{V}_{\text{out}}$ denote the modality-indicator and task-output vocabularies, respectively, and $\mathbb{I}(\cdot)$ is the indicator function. 

\subsection{Output Representations}
\label{sec:output_rep}
Following Section~\ref{sec:model_arch}, we instantiate the target sequence $\mathbf{y}$ for SD and auxiliary tasks.
We compare two output representations: \texttt{event}-based and \texttt{frame}-based. For the \texttt{event}-based representation, we use a representative timestamp-tuple serialization and keep it fixed across experiments, as minor serialization variants showed the same qualitative trends in preliminary runs. The \texttt{frame}-based representation predicts fixed-resolution speaker activity, following the standard formulation used by many SD systems, and serves as a controlled contrast to the compact event-level generation format. While frame-level outputs are natural for models that directly predict speaker activity over time, they lead to longer and more repetitive target sequences for an autoregressive SpeechLM.

Full-meeting recordings are divided into 30-second segments with a 10-second stride during training, and decoded as non-overlapping 30-second segments during inference. Reference speakers are assigned recording-level IDs by first appearance and kept consistent within each recording. Since inference is performed independently on each segment, the generated speaker symbols denote local anonymous speaker hypotheses rather than recording-level identities. For example, a local \texttt{<spk1>} in one segment is not guaranteed to correspond to \texttt{<spk1>} in another segment without additional information. Full-meeting consistency therefore requires an explicit speaker-linking or speaker-tracking mechanism, which we analyze in Section~\ref{sec:spk_linking}.


\subsubsection{\texttt{Event}-based representation} Each speaker turn is encoded as a five-token tuple consisting of the speaker ID, a begin-of-time marker, the starting timestamp, the ending timestamp, and an end-of-time marker. Timestamps are quantized at 0.1-second resolution, with each discrete time point represented by a dedicated timestamp token. e.g., 
\vspace{-0.1cm}
\[\small
\mathbf{y}_{sd}^{event}=[\texttt{<spk1> \space <bot> \space  <0.0> \space  <2.0> \space <eot>}]
\vspace{-0.1cm}
\]

\noindent $\mathbf{y}_{sd}^{event}$ is an ordered token sequence and shows that \texttt{<spk1>} is speaking for the first two seconds. If no speaker is active, we represent the output sequence as
\vspace{-0.1cm}
\[\small
\mathbf{y}_{sil}^{event}=[\texttt{<sil> \space <bot> \space  <0.0> \space  <2.0> \space <eot>]}
\vspace{-0.1cm}
\]

\subsubsection{\texttt{Frame}-based representation} We output one token per 0.1 second. For example, over the first 0.5 seconds, if \texttt{<spk1>} is active during the 0.1–0.3s interval, the corresponding representation is
\vspace{-0.1cm}
\[\small
\mathbf{y}_{sd}^{frame}=[\texttt{<sil> <spk1> <spk1> <sil> <sil>}]
\vspace{-0.1cm}
\] 
For overlapping speakers, we explicitly model up to three speakers (e.g., \texttt{<overlap\_spk\_1\_2\_3>}). When more than three speakers talk simultaneously, a rare case in conversational datasets, we map the region to a single token, \texttt{<overlap\_spk\_4\_more>}.
For a 30-second audio input, the model generates 300 frame-level tokens. For such long outputs, we add transition markers to help autoregressive SpeechLMs better align with speaker changes by indicating where turns should end.
An example sequence is
\vspace{-0.1cm}
\[\small
\begin{aligned}
    {\mathbf{y}'}_{sd}^{frame} &=[\texttt{<sil> <sc\_start> <spk1> <spk1>} \\
    &  \texttt{<sc\_end> <sil> <sil>]}
\end{aligned}
\vspace{-0.1cm}
\]
Each speaker-turn segment is augmented with explicit transition tokens, \texttt{<sc\_start>} and \texttt{<sc\_end>}, to mark its beginning and end.
Table~\ref{tab:ami_token_length} reports the average and standard deviation of token counts for AMI under the \texttt{event}- and \texttt{frame}-based representations.

\begin{table}[t]
	\centering
	\setlength{\tabcolsep}{3pt}
	\renewcommand{\arraystretch}{1.25}
    \caption{Average token lengths (mean $\pm$ standard deviation) for the SD, SAD, OD, and STC tasks on AMI under \texttt{event}- and \texttt{frame}-based representations. The \texttt{frame}-based setup includes transition markers.}

		\begin{tabular}{l|c|c|c|c}
			\hline
			\textbf{Format} & \textbf{SD} & \textbf{SAD} & \textbf{OD} & \textbf{STC} \\
			\hline
			\texttt{event} & 40.0 $\pm$ 21.0 & 18.8 $\pm$ 10.5 & 18.2 $\pm$ 13.3 & 6.7 $\pm$ 2.1 \\
			\texttt{frame} & 323.2 $\pm$ 15.7 & 307.6 $\pm$ 8.0 & 307.4 $\pm$ 8.7 & 6.7 $\pm$ 2.1 \\
			\hline
		\end{tabular}
    \vspace{-0.4cm}
	\label{tab:ami_token_length}
\end{table}

\subsubsection{Auxiliary Tasks} In addition to SD prediction, we introduce three auxiliary tasks: SAD, OD, and STC. SAD identifies speech regions within the audio, OD detects overlapping speech segments, and STC counts the number of single and overlapped speaker turns. The SAD and OD tasks can be expressed in either the \texttt{event}- or \texttt{frame}-based representation in a similar manner as SD. For STC, we instead use a single unified format: 
\vspace{-0.1cm}
\[\small
\begin{aligned}
    \mathbf{y}_{stc}&=[\texttt{<spk1\_count> <1\_count> <spk2\_count>} \\
    \texttt{<1\_count>}]
\end{aligned}
\vspace{-0.1cm}
\]

\noindent indicating that \texttt{spk1} and \texttt{spk2} each produce one speaker turn.
Due to the long-tail distribution of STC, we cap the STC value at ten, grouping speakers with ten or more turns into a single \texttt{10\_count} token. This token accounts for less than 2\% of the corpus and avoids token sparsity.

\subsubsection{Single-Stream and Multi-Stream Organizations} 

We explore two ways of organizing $\mathbf{y}$ when auxiliary tasks are included. In the single-stream organization, illustrated in Figure~\ref{fig:model} all task outputs are concatenated into one ordered sequence. For example, when SAD, OD, and SD are included, the target output sequence can be written as:
\[
\small
\begin{aligned}
\mathbf{y}
=
[&\texttt{<start\_sad>}, \mathbf{y}_{\mathrm{sad}}, \texttt{<end\_sad>}, \\
 &\texttt{<start\_od>}, \mathbf{y}_{\mathrm{od}}, \texttt{<end\_od>}, \\
 &\texttt{<start\_sd>}, \mathbf{y}_{\mathrm{sd}}, \texttt{<end\_sd>}].
\end{aligned}
\]
In the multi-stream organization, task outputs are assigned to separate output streams:
\[
\small
\begin{aligned}
\mathbf{y}^{(1)} &= [\texttt{<start\_sd>}, \mathbf{y}_{\mathrm{sd}}, \texttt{<end\_sd>}], \\
\mathbf{y}^{(2)} &= [\texttt{<start\_od>}, \mathbf{y}_{\mathrm{od}}, \texttt{<end\_od>}], \\
\mathbf{y}^{(3)} &= [\texttt{<start\_sad>}, \mathbf{y}_{\mathrm{sad}}, \texttt{<end\_sad>}].
\end{aligned}
\]
The single-stream organization preserves a stepwise autoregressive structure in which auxiliary tasks can guide the final SD output, while the multi-stream organization reduces the effective output length by generating task outputs in parallel.

We use different task-output organizations for the two SD output representations. For the \texttt{event}-based representation, timestamped tuples are already compact, so the single-stream organization remains tractable even with multiple auxiliary tasks. We also explored a multi-stream organization in which SAD, OD, and SD are assigned to separate output streams, but this substantially degrades performance, likely because the model must handle heterogeneous output formats across streams. For the \texttt{frame}-based representation, each task requires a full frame-level sequence, making the single-stream organization much longer and more prone to format drift. We therefore use the single-stream organization for the \texttt{event}-based representation and the multi-stream organization for the \texttt{frame}-based representation, which keeps the frame-level output length roughly bounded by the 300 frame steps in a 30-second segment.

\section{Experiments}
\subsection{Experimental Setting}

We fine-tune a 1.7B-parameter decoder-only Transformer with an autoregressive delay LM head~\cite{copet2023simple}. Training uses token-based batching, mixed precision, SpecAugment~\cite{park19e_interspeech} with time masking, and AdamW~\cite{loshchilov2018decoupled} (learning rate 3e-4, weight decay 1e-6) with gradient clipping at 1.0 and a 4k-step warm-up. Those hyperparameters are selected on AMI and reused for AISHELL-4 and AliMeeting. We fine-tune the full SpeechLM model for 10 epochs on one NVIDIA H100 GPU and average the three best checkpoints based on development token accuracy. For domain adaptation, SpeechLM is first trained on simulated data for 25 epochs and then fine-tuned for 3 epochs under either dataset-specific or joint training across AMI, AISHELL-4, and AliMeeting, using one NVIDIA A100 GPU. We use AdamW with learning rates of (1$\mathrm{e}{-4}$) and (2$\mathrm{e}{-4}$) for dataset-specific and joint adaptation, respectively. We evaluate all systems using DER, including miss detection (MISS), false alarm (FA), and speaker confusion (SC), with zero collar tolerance and score overlapped speech regions. For auxiliary-task inspection, we compute SAD and OD F1 scores on the 0.1-second frame grid, treating SAD as speech/non-speech detection and OD as overlap/non-overlap detection.

\subsection{Post-processing}
\label{sec:spk_linking}

We use greedy decoding and stop when the task-specific end marker is generated, e.g., \texttt{<end\_sd>} for SD. To extract SD outputs, we identify the sequence enclosed by \texttt{<start\_sd>} and \texttt{<end\_sd>}. For \texttt{event}-based outputs, we parse five-token timestamped tuples; for \texttt{frame}-based outputs, we retain only silence and speaker-activity tokens and ignore transition markers. Outputs longer than the input duration are truncated, while shorter outputs are padded with silence. 
When \texttt{frame}-based outputs contain parsing errors due to format drift in longer, marker-rich sequences, we retain only valid task-associated tokens.
We also cross-check SD predictions with SAD predictions for speech-activity consistency and apply an 11-frame median filter after converting outputs to frame-level speaker activities.

\subsection{Speaker-Linking Diagnostics}
To analyze speaker linking across 30-second segments, we evaluate four post-processing settings. The first three settings are used as diagnostic comparisons on the best-performing \texttt{event}-based configuration with SAD$\rightarrow$OD$\rightarrow$SD setting (see Table~\ref{tab:DER_inference}), while the fourth setting is used as our default post-processing backend because it achieves the best overall DER. Given a recording waveform $\mathbf{x}$, let $\hat{\mathbf{y}}_{sd}$ denote the decoded output sequence after adding each window offset to the decoded timestamps. We parse $\hat{\mathbf{y}}_{sd}$ into a SD hypothesis
\begin{equation}
    \hat{\mathcal{H}}
    =
    \{(\hat{b}_i,\hat{e}_i,\hat{s}_i)\}_{i=1}^{I},
\end{equation}
where $I$ is the number of predicted speaker-labeled segments, and $(\hat{b}_i,\hat{e}_i,\hat{s}_i)$ denotes the start time, end time, and local SpeechLM speaker symbol of segment $i$.
\subsubsection{Raw SpeechLM output}
The first setting uses the raw decoded SpeechLM output $\hat{\mathcal{H}}$ without additional linking. 
\subsubsection{Oracle within-window relabeling} The second setting applies oracle within-window speaker permutation. For each decoded 30-second window $w$, we find the one-to-one mapping $\pi_w$ from local SpeechLM speaker symbols to reference speakers that maximizes total temporal overlap:
\begin{equation}
    \pi_w^*
    =
    \arg\max_{\pi_w}
    \sum_{i \in w}
    \mathrm{overlap}
    \left(
    [\hat{b}_i,\hat{e}_i],
    \mathcal{R}_{\pi_w(\hat{s}_i)}
    \right),
\end{equation}

where $\mathcal{R}_{k}$ denotes the reference activity regions of speaker $k$. We then relabel each predicted segment by replacing $\hat{s}_i$ with $\pi_w^*(\hat{s}_i)$ and concatenate the relabeled windows into a full-meeting hypothesis before final evaluation. This setting is not deployable in practice because it uses oracle reference timestamps and speaker identities, but serves as a diagnostic upper bound that removes cross-window speaker permutation errors without changing predicted speech boundaries.

\subsubsection{WeSpeaker enrollment relabeling} 

We construct an oracle enrollment inventory using ResNet34-LM2 WeSpeaker embeddings~\cite{Wang2024}. For each reference speaker $k$, we select a clean non-overlapping enrollment segment $a_k$ of at most five seconds and compute a length-normalized embedding $\bar{\mathbf{e}}_k$. For each predicted segment $\hat{\mathcal{H}}_i=\{(\hat{b}_i,\hat{e}_i,\hat{s}_i)\}$, we compute an embedding $\bar{\mathbf{z}}_i$ from the corresponding waveform region:
\begin{equation}
\small
    \bar{\mathbf{e}}_k =
    \frac{\mathrm{WeSpeaker}(a_k)}
         {\|\mathrm{WeSpeaker}(a_k)\|_2},
    \quad
    \bar{\mathbf{z}}_i =
    \frac{\mathrm{WeSpeaker}(\mathbf{x}[\hat{b}_i:\hat{e}_i])}
         {\|\mathrm{WeSpeaker}(\mathbf{x}[\hat{b}_i:\hat{e}_i])\|_2}.
\end{equation}
Let $\mathcal{E}=(\bar{\mathbf{e}}_1,\ldots,\bar{\mathbf{e}}_K)$, where $K\leq5$ is the number of reference speakers. We replace only the raw SpeechLM speaker symbol with the closest enrolled speaker,
$\hat{k}_i=\arg\max_{k\in\{1,\ldots,K\}}\bar{\mathbf{z}}_i^\top\bar{\mathbf{e}}_k$.
This measures how much SC error can be reduced using clean oracle enrollment audio.

\subsubsection{VBx speaker linking} The fourth setting applies VBx post-processing following the Pyannote inference recipe~\cite{han2025leveraging, Bredin2023}. SpeechLM predictions provide the temporal speaker-activity hypotheses, while VBx links speaker identities across segments using ResNet34-LM2 WeSpeaker embeddings~\cite{Wang2024} and VBx clustering~\cite{landini2022bayesian,Bredin2023}.


\section{Results}

\begin{table}[!t]
	\centering
    \caption{
		DER (\%) on AMI for different training sequences incorporating
		various auxiliary tasks using the \texttt{event}-based representation with VBx post-processing. The best DER result is \textbf{bolded}.
	}
	\renewcommand{\arraystretch}{1.2}
		\begin{tabular}{l c | l c}
			\hline
			\textbf{Setting} & \textbf{DER} &
			\textbf{Setting} & \textbf{DER} \\
			\hline
			
			SD only & 26.48 &
			\textbf{SAD $\rightarrow$ OD $\rightarrow$ SD} & \textbf{24.40} \\
			\cline{1-2}
			
			SD $\rightarrow$ SAD & 26.22 &
			SAD $\rightarrow$ SD $\rightarrow$ STC & 25.74 \\
			
			SAD $\rightarrow$ SD & 25.69 &
			SAD $\rightarrow$ SD $\rightarrow$ OD & 25.50 \\
			
			SD $\rightarrow$ OD & 25.39 &
			OD $\rightarrow$ SAD $\rightarrow$ SD & 25.33 \\
            \cline{3-4}

			OD $\rightarrow$ SD & 26.25 &
			SAD $\rightarrow$ OD $\rightarrow$ SD $\rightarrow$ STC & 25.46 \\
			
			SD $\rightarrow$ STC & 25.79 &
			OD $\rightarrow$ SAD $\rightarrow$ SD $\rightarrow$ STC & 25.48 \\
			
			STC $\rightarrow$ SD & 29.16 &
			SAD $\rightarrow$ SD $\rightarrow$ OD $\rightarrow$ STC & 25.68 \\
			
			& &
			SAD $\rightarrow$ SD $\rightarrow$ STC $\rightarrow$ OD & 26.12 \\
			\hline
		\end{tabular}

	\vspace{-0.4cm}
	\label{tab:der_event_sequence}
\end{table}

\begin{table}
    \centering
    \setlength{\tabcolsep}{4pt}
    \renewcommand{\arraystretch}{1.12}
    \caption{DER (\%) decomposition on AMI under the SAD$\rightarrow$OD$\rightarrow$SD setting. The upper panel compares different inference settings, while the lower panel further decomposes the VBx post-processed output into single-speaker and overlapped-speaker regions.}
    \label{tab:DER_inference}
    \begin{tabular}{l|cccc}
        \hline
        \textbf{Setting} & \textbf{MISS} & \textbf{FA} & \textbf{SC} & \textbf{DER} \\
        \hline
        \multicolumn{5}{l}{\textit{Whole recording}} \\
        SpeechLM raw        
        & 13.12 & 4.45 & 48.73 & 66.30 \\
        SpeechLM + oracle relabeling  & 13.12 & 4.42 & 16.58 & 34.12 \\
        SpeechLM + WeSpeaker relabeling & 12.44 & 4.69 & 11.55 & 28.67 \\
        SpeechLM + VBx speaker linking     
        & 13.11 & 4.43 & 6.86  & 24.40 \\
        \hline
        \multicolumn{5}{l}{\textit{Region-wise breakdown of VBx post-processing}} \\
        Single-speaker regions 
        & 4.39  & 1.89 & 5.99 & 12.26 \\
        Overlap regions    
        & 36.68 & 0.18 & 9.15 & 46.01 \\
        \hline
    \end{tabular}
    \vspace{-0.4cm}
\end{table}

\subsection{Effects of Auxiliary-Task Ordering}
Inspired by chain-of-thought prompting in LLMs~\cite{wei2022chain}, we examine how the ordering of auxiliary tasks, SAD, OD, and STC, affects autoregressive decoding and SD performance. Table~\ref{tab:der_event_sequence} shows results for the \texttt{event}-based representation on AMI. Coarse-to-fine auxiliary-task chains consistently outperform the SD-only baseline of 26.48\% DER, confirming the value of auxiliary-task conditioning. Among two-stage chains, SAD$\rightarrow$SD performs better than SD$\rightarrow$SAD, indicating that SAD cues help guide later speaker attribution. OD and STC are more effective when placed after SD, where they provide complementary higher-level information. The best three-stage chain, SAD$\rightarrow$OD$\rightarrow$SD, achieves 24.40\% DER, suggesting that coarse temporal cues should precede speaker prediction. Auxiliary-task results further follow the expected difficulty hierarchy, with SAD reaching about 91\% F1 and OD about 78\% F1 on AMI. Adding STC as a fourth task yields little additional improvement, indicating diminishing returns beyond three-task chains.

\begin{table}[!t]
	\centering
    \caption{DER (\%) for different training settings on AMI under \texttt{event}- and \texttt{frame}-based representations with VBx post-processing. 
    Best DER results are \textbf{bolded}.}
		\begin{tabular}{l|cc}
			\hline
			\textbf{Training Setting} & \texttt{event} & \texttt{frame} \\
			\hline
			
			SD only
			& 26.48 & 37.88 \\
			\hline
			
			SAD $\rightarrow$ SD
			& 25.69 & \textbf{29.37} \\
			
			SD $\rightarrow$ OD
			& 25.39 & 56.66 \\
			
			SD $\rightarrow$ STC
			& 25.79 & 41.22 \\
			\hline
			
			SAD $\rightarrow$ OD $\rightarrow$ SD
			& \textbf{24.40} & 56.94 \\
			\hline
			
			SAD $\rightarrow$ OD $\rightarrow$ SD $\rightarrow$ STC
			& 25.46 & 35.96 \\
			\hline
			
		\end{tabular}
    \vspace{-0.3cm}
\label{tab:der_event_frame}
\end{table}

\begin{table}[!t]
\centering
\caption{DER (\%) on AMI, AISHELL-4, AliMeeting, and their cross-dataset average for data-specific and joint training, with and without adaptation, using the \texttt{event}-based SAD$\rightarrow$OD$\rightarrow$SD representation with VBx post-processing. Best DER results under each configuration are shown in \textbf{bold}.
}
\setlength{\tabcolsep}{2pt}
\renewcommand{\arraystretch}{1.1}
\begin{tabular}{l | c c c | c}
\hline
Training Method
& AMI & AISHELL-4 & AliMeeting & Avg\\
\hline
Data-specific
& 24.40 & 13.75 & 29.60 & 22.58 \\
\quad + adaptation
& 23.00 & \textbf{12.48} & 23.00 & 19.50 \\
\hline
Joint
& 24.54 & 14.14 & 28.58 & 22.42 \\
\quad + adaptation
& \textbf{22.08} & 13.29 & \textbf{22.81} & \textbf{19.40} \\
\hline
\end{tabular}

\vspace{-0.3cm}
\label{tab:training_comparison_full}
\end{table}

\begin{table*}[t]
\centering
\caption{Comparison of DER (\%) between SpeechLM-based diarization models and representative prior systems at full-meeting level in the literature. $^\dagger$G-STAR reports DER but does not specify the AMI microphone/mixture condition.}

\begin{tabular}{l|c|ccc}
\hline
\textbf{Model} & \textbf{Simulated Data Used} & \textbf{AMI} & \textbf{AISHELL-4} & \textbf{AliMeeting} \\
\hline
\multicolumn{5}{l}{\textbf{Clustering-based}} \\
\hline
Transcribe-to-Diarize~\cite{kanda2022transcribe} & -- & 24.4 & -- & -- \\
VAD+VBx+OSD~\cite{landini2024diaper} & -- & 22.4 & 15.8 & 28.8 \\
pyannote 3.1~\cite{Plaquet23} & -- & 18.8 & 12.2 & 24.4 \\
\hline
\multicolumn{5}{l}{\textbf{EEND-based}} \\
\hline
SA-EEND~\cite{horiguchi2022encoder} & 100k mixtures on 2 speakers, $\sim$2433 h & 27.7 & -- & -- \\
EEND-EDA~\cite{Chen2024} & 100k mixtures on 1--5 speakers & 21.6 & -- & -- \\
EEND-VC~\cite{palka2025vbx} & -- & 14.2 & 9.8 & 12.4 \\
EEND-TA + FT~\cite{broughton2025pushing} & $>$80k h, 1--8-spk mixtures & 11.0 & 12.2 & 11.4 \\
\hline
\multicolumn{5}{l}{\textbf{Token-based}} \\
\hline
G-STAR$^\dagger$~\cite{peng2026gstar} & -- & 32.2 & -- & -- \\
SLIDAR~\cite{cornell2024one} & 5000 h & 27.1 & -- & -- \\
SpeechLM (data-specific) & 3304 h & 23.0 & 12.48 & 23.0 \\
SpeechLM (joint) & 3304 h & 22.1 & 13.29 & 22.81 \\
\hline
\end{tabular}
\label{tab:der_other_models}
\vspace{-0.4cm}
\end{table*}

\subsection{Speaker Linking and Overlap Error Analysis}

Building on the speaker-linking settings in Section~\ref{sec:spk_linking}, Table~\ref{tab:DER_inference} separates the temporal speaker-activity structure generated by SpeechLM from the recording-level speaker consistency supplied by explicit linking. The major limitation of raw SpeechLM decoding is SC rather than FA or MISS errors.
Under the best \texttt{event}-based representation, raw decoding obtains 66.30\% DER, with SC contributing 48.73\%. 
Oracle within-window speaker relabeling reduces SC from 48.73\% to 16.58\% and DER from 66.30\% to 34.12\%, showing that much of the error comes from cross-window speaker-token inconsistency. The remaining SC, however, suggests that local SpeechLM speaker symbols are still imperfect, even after oracle permutation.
Relabeling the generated speaker labels with WeSpeaker embeddings extracted from clean enrollment audio further reduces SC to 11.55\%. The slight MISS/FA changes are due to independent segment-level relabeling: multiple local SpeechLM speaker symbols may collapse to the same enrolled speaker, causing overlapping same-speaker regions to be merged. VBx post-processing achieves the best overall DER of 24.40\% and further reduces SC to 6.86\%. However, the region-wise decomposition reveals that this improvement is concentrated in single-speaker regions, where DER is 12.26\%; overlapped regions remain difficult, with 46.01\% DER driven primarily by missed speech. This indicates that explicit speaker linking can substantially mitigate identity errors, while overlap handling remains a major limitation.

\vspace{-0.2cm}
\subsection{\texttt{Event}- vs. \texttt{Frame}-based Representations}

Table~\ref{tab:der_event_frame} compares \texttt{event}- and \texttt{frame}-based representations on AMI using the best configuration for each task-chain length from Table~\ref{tab:der_event_sequence}. The \texttt{event}-based representation consistently outperforms the \texttt{frame}-based alternative and is substantially more stable across auxiliary-task configurations, with DER remaining within a narrow range of 24.40--26.48\%. In contrast, the \texttt{frame}-based representation is more sensitive to the output sequence design: adding SAD improves performance from 37.88\% to 29.37\%, but OD-related chains lead to large degradations, reaching over 56\% DER. This suggests that frame-level supervision can provide useful speech-activity cues, but long frame-level outputs are brittle for autoregressive decoding, especially when multiple temporally dense auxiliary tasks must be generated. Overall, the compact \texttt{event}-based representation provides a more reliable format for SpeechLM-based meeting diarization.


\subsection{Effect of Adaptation}

To investigate the effect of domain adaptation, we first train the SpeechLM-based SD system on large-scale simulated mixtures using the three-stage chain SAD$\rightarrow$OD$\rightarrow$SD. We then adapt the model to all three meeting datasets. We compare data-specific training and joint training across all three datasets, both before and after adaptation.
As shown in Table~\ref{tab:training_comparison_full}, adaptation consistently improves DER across all datasets and training strategies. In particular, AliMeeting benefits the most from adaptation, with DER reduced from 28.58\% to 22.81\% under joint training. Data-specific adaptation yields larger gains on AISHELL-4, while joint training with adaptation achieves the best overall average DER, indicating robust cross-dataset generalization without dataset-specific adaptation.

\subsection{Comparison with Previous Models}
Table~\ref{tab:der_other_models} compares our SpeechLM-based systems with representative SD systems grouped by modeling paradigm. We only include prior systems that report DER performance at the full-meeting level. Since these systems differ in microphone conditions, training-data scale, architectures, and post-processing, the comparison should be interpreted as a broad reference rather than a strictly controlled benchmark. Specialized clustering-based and EEND-based systems generally remain strongest in absolute DER. Among token-based approaches with clustering-based speaker-linking backends, SpeechLM achieves lower DER than SLIDAR~\cite{cornell2024one} on AMI while using less simulated training data, and is evaluated consistently across AMI, AISHELL-4, and AliMeeting. 
G-STAR~\cite{peng2026gstar} uses a cache-conditioned Sortformer-style tracker to maintain persistent speaker states, further highlighting the need for explicit speaker tracking in token-based SD. Despite differences in formulation, inference protocol, and microphone condition, its meeting-level results are consistent with our finding that robust full-meeting SD requires speaker linking or tracking beyond raw autoregressive speaker symbols.

\section{Conclusion}

This work studies autoregressive SpeechLMs for full-meeting SD. Event-based outputs are more compact and robust than frame-based generation, and coarse-to-fine auxiliary prediction (SAD$\rightarrow$OD$\rightarrow$SD) improves DER. However, raw SpeechLM speaker symbols do not preserve recording-level identity across independently decoded segments; explicit speaker linking substantially reduces speaker confusion error, while overlap-related misses remain a major error source. Thus, SpeechLMs provide useful temporal diarization structure, but robust full-meeting SD still requires persistent speaker tracking and stronger overlap modeling. 


\section*{Acknowledgement}
This work used the Bridges2 system at PSC and Delta system at NCSA through allocation CIS210014 from the Advanced Cyberinfrastructure Coordination Ecosystem: Services \& Support (ACCESS) program, supported by National Science Foundation grants \#2138259, \#2138286, \#2138307, \#2137603, and \#2138296.

\section*{Generative AI Use Disclosure}

The authors used ChatGPT to assist with language editing, wording refinement, and LaTeX formatting in portions of the Abstract, Introduction, Methods, Results discussion, table captions, and related explanatory text. All technical content, experimental design, analysis, results, and conclusions were developed, reviewed, and verified by the authors.

\bibliographystyle{IEEEtran}
\bibliography{refs}

\end{document}